\documentclass[twocolumn,aps,prb,10pt,amsmath,amssymb,groupedaddress,nofootinbib,]{revtex4-2}

\usepackage{graphicx}
\usepackage{dcolumn}
\usepackage{bm}
\usepackage[dvipsnames]{xcolor}
\usepackage{tabu}
\usepackage{xcolor}
\usepackage{hyperref}
\usepackage{outlines}
\usepackage{multirow}
\usepackage{url}
\usepackage{physics}
\usepackage{pifont}

\begin{document}
\title{
Oxygen deficiency and valency reconstruction in multiferroic V-doped HfO$_2$}
\author{Vincenzo Fiorentini}
\affiliation{Department of  Physics, University of Cagliari, Cittadella Universitaria, I-09042 Monserrato (CA), Italy}
\affiliation{Institute for Materials Science and Max Bergmann Center for Biomaterials, TU Dresden, D-01062 Dresden, Germany}
\date{\today}

\begin{abstract}
%
The interplay of oxygen deficiency and vanadium multiple valency in the candidate multiferroic V-doped $Pca2_1$ hafnia HfO$_2$ is studied by first-principles calculations. Low-lying V majority gap states accept electrons from oxygen-vacancy donors, reducing their formation energy, and converting  nominal V$^{4+}$ centers into V$^{3+}$. The resulting local magnetization and screening changes are reflected in the calculated V core-level shifts, which are consistent with the experimentally observed XPS signatures. The calculated  V$^{3+}$/V$^{4+}$ population ratio determined by oxygen vacancies only matches experiment in reducing conditions, suggesting that additional electron reservoirs may contribute under ALD growth conditions. A similar scenario also seems to apply to the recently observed multiferroicity in Cr-doped hafnia, where oxygen deficiency is expected from the synthesis route.
\end{abstract}

\maketitle

\section{Introduction}

Multiferroic materials, exhibiting ferroelectricity and ferromagnetism, are rare. Recently \cite{io} a promising route to multiferroicity was proposed, via doping of ferroelectric hafnia (orthorhombic $Pca2_1$ HfO$_2$) with vanadium (V). Ferroelectric V:HfO$_2$ was indeed produced in Ref.\cite{epfl}, though magnetism was not studied;  multiferroicity in the analogous system Cr:HfO$_2$ was also recently observed \cite{cr}. 

As a substitution for Hf in otherwise perfect hafnia (as in Ref.\onlinecite{io}) V is by construction a nominal 4+ ion with one electron in excess over Hf. Experiment \cite{epfl} showed that V adopts both 3+ and  4+ valency, with a population ratio of about 2.8 around 6\% concentration \cite{stima}. Inspired by this observation, in this paper I discuss the multiple valency problem via ab initio calculations of V in HfO$_2$ at different concentrations in the presence of oxygen vacancies. The basic idea is that V has multiple valency, i.e. it can deploy different numbers of electrons in different environments, and it can act as a 2+, 3+, 4+, and 5+ ion, forming for example V$_2$O$_3$,  VO$_2$, and V$_2$O$_5$ for increasing oxygen availability. 

Intuitively, V embedded in hafnia --whose Hf cations act unflinchingly as 4+ ions-- could adopt a 3+ state capturing electrons from donors, for example O vacancies; reciprocally, the presence of V could make those costly defects affordable. As discussed below, the electronic structure of substitutional V and O vacancies is indeed conducive to this sort of energetic cooperation. Naturally, the nominally 3+ and 4+ states  need to be characterized, and it so happens that core level shifts and local magnetization naturally allow to identify them unambiguously.
The results also shed  light on the recently observed multiferrocity in hafnia \cite{cr} doped with Cr, which neighbors V in the Periodic Table.

\section{Methods}

As in Ref.\onlinecite{io},  observables are obtained from first principles within density functional theory for multiple configurations of V substituting for Hf in ferroelectric HfO$_2$, together with oxygen vacancies. To avoid confusion, the oxygen vacancy is denoted by X in this paper. The VASP code (v.6.5) \cite{vasp} and the PAW formalism \cite{PAW} are used, with  the datasets {\tt V}, {\tt Hf} and {\tt O$_{\tt s}$} at a cutoff of 350 eV (1.3 times the maximum recommended cutoff).  The density-functional calculations are spin-polarized GGA+U, generalized gradient  PBE \cite{PBE} functional and Hubbard  U corrections in the Dudarev version \cite{LDAU}. The Hubbard correction is U--J=3 eV on V $d$ states only, a  standard \cite{V-U} setting validated \cite{io}  by comparison with hybrid functional calculations \cite{HSE}.

A 96-atom 2$\times$2$\times$2 replica of the $Pca2_1$  primitive cell is used, with supercell lattice vectors vs concentration as calculated in \cite{io}. Internal geometries are optimized in all cases with typical force threshold 0.02 eV/\AA,  a 2$\times$2$\times$2 k-space grid, and a Gaussian smearing of 0.02 eV. Starting from  the lowest-energy V configurations of Ref.\onlinecite{io}, I study the energetics of formation of one or two oxygen vacancies X per supercell and different numbers of V, and several configurations are explored to search for  clustering effects driven by X. 

\section{Results}
\subsection{Schematics of V$_n$X coupling}
\label{gencons}

All the considerations in Sec.II of Ref.\onlinecite{io} on V:HfO$_2$ remain valid for the present system.
In particular, the majority $t_{2g}$ states are split into non-degenerate states already in undoped hafnia, and  hence a fortiori in the presence of  disorder by random substitutions and vacancy creation. As shown in Ref.\onlinecite{io}, occupied gap states induced by single and multiple V$_n$ substitutions lie generally in the bottom 0.5-1 eV region of the hafnia gap; unoccupied majority states live predominantly around midgap, so 2-2.5 eV above the valence band edge.   

\begin{figure}[ht]
\centerline{\includegraphics[width=1\linewidth]{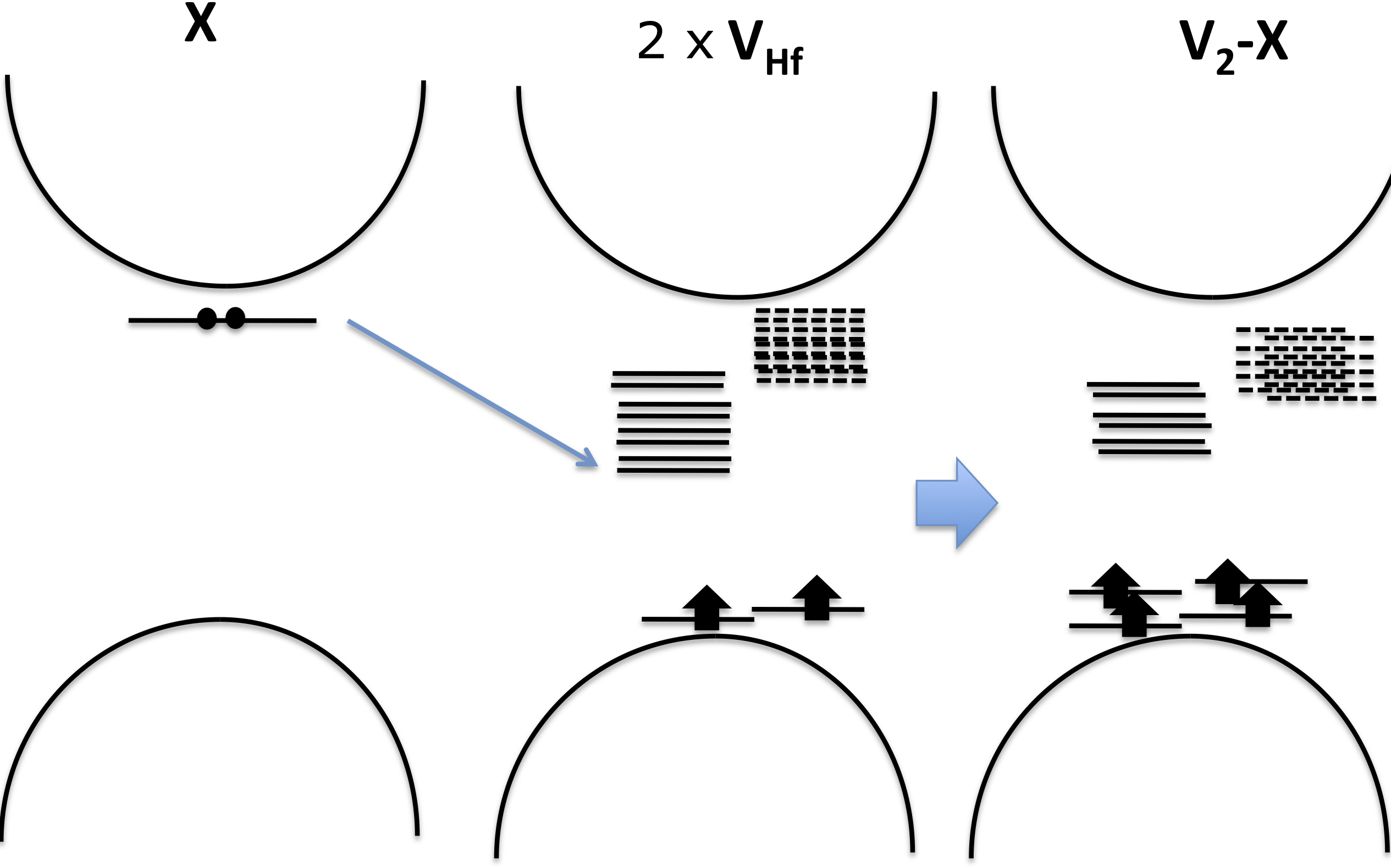}}
\caption{\label{schema} Schematic of electron transfer in the V$_2$ X system.}
\end{figure}

As in many other oxides, the oxygen vacancy X is a double donor with deep(ish) states around 1 eV from the conduction edge. Given the large gap of hafnia (5.9 eV experimentally and 4.6 eV in GGA+U), V$_n$ states are appreciably lower than X-induced states. One therefore expects that V$_n$  will accept electrons from X, changing occupation, magnetization, and effective valency of one or more of its component V. This donor-acceptor electronic rearrangement  leads to a typical 1.5-2 eV energy gain per electron, i.e. conservatively a 3-3.5 eV gain per oxygen vacancy formed (plus  ionic relaxation or clustering energy gains).  Figure \ref{schema} exemplifies schematically the process for V$_2$ X (though the Figure is not drawn to scale, the number and approximate positions of energy levels are correctly described).

\begin{figure}[ht]
\centerline{\includegraphics[width=1\linewidth]{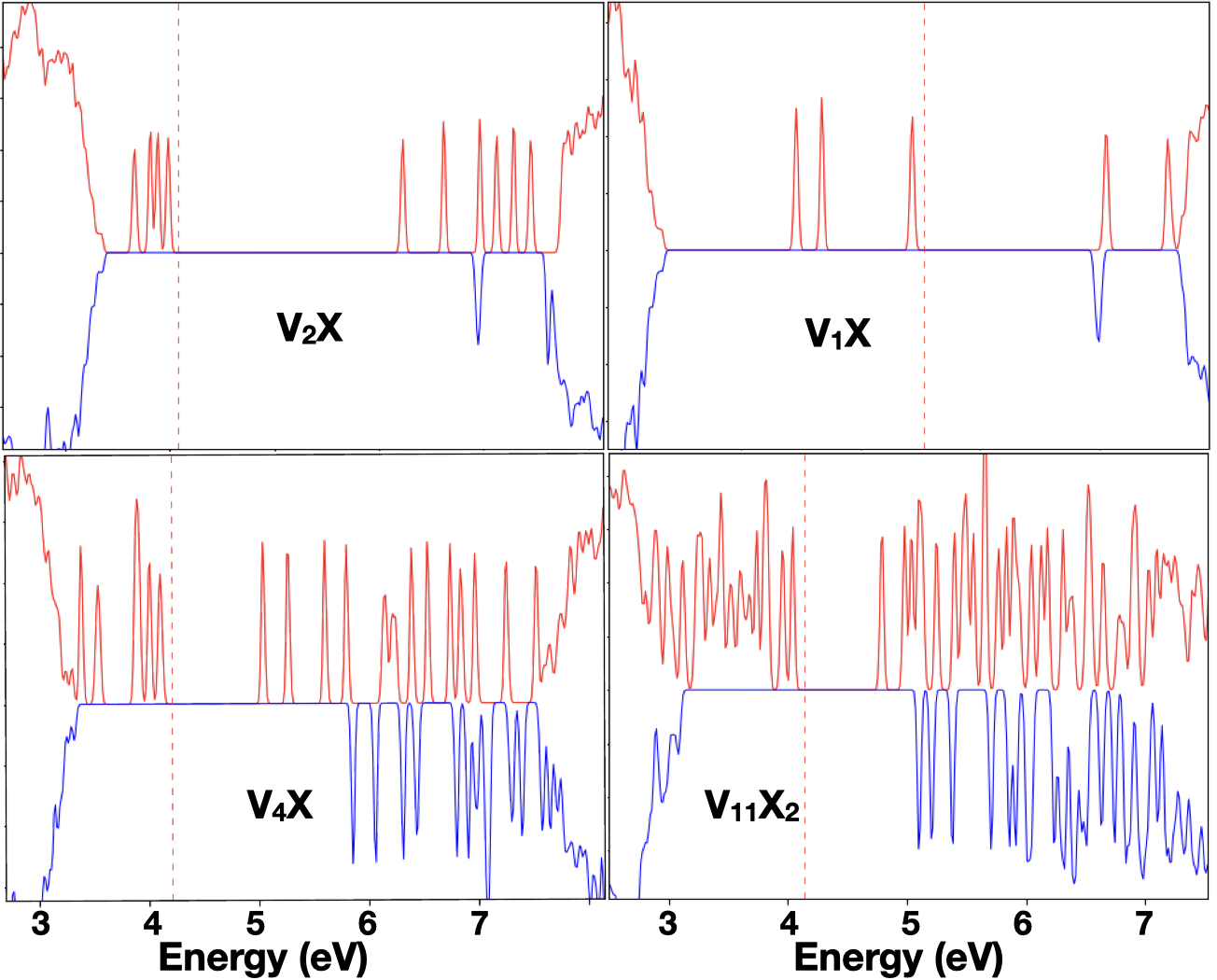}}
\caption{\label{dos} Density of states (arbitrary units) of V$_n$X$_k$ for ($n$,$k$)=(1,1), (2,1), (4,1), and (11,2), showing the presence of peaks integrating to $n$+2$k$ states below the Fermi level (vertical dashed lines).}
\end{figure}

The magnetization  M(V$_n$)=$n$ $\mu_B$ of V$_n$ is obviously affected by the presence of X. As it  is  practically always the case that spins are ferromagnetically aligned (as in the Figure), the magnetization  becomes 
M(V$_n$X)=$n$+2$k$ $\mu_B$.
This corresponds to $n$+2$k$ peaks (more precisely, peaks integrating to $n$+2$k$ electrons) in the occupied V$_n$X$_k$ density of states below the Fermi energy in the lower part of the hafnia gap, as shown in Figure \ref{dos}. This behavior is typical, though  occasionally a minority state drops below the Fermi level, reducing the magnetization. 

\subsection{Nominal valency and core levels}

Since X  releases two electrons to lower-lying V states, the magnetization of two V's in the system will change from 1 $\mu_B$ to 2 $\mu_B$. Because the magnetization is calculated as an integral inside a small V-centered sphere, one can plausibly relate the magnetization and effective valency of each specific V atom. The 4+ state of V has M=1 \cite{io}; a V with M=2 has gained one electron, and can be  labeled 3+; a V acquiring 2 electrons and having M=3 is labeled 2+; a V losing its native electron has M=0 and valency 5+. Again, this connection is reasonable because the magnetization change is localized. Such being the case, the local screening change should be mirrored in the core level shifts observed by X-ray photoelectron spectroscopy (XPS). 
 
 The repertoire of XPS  indicates that, assuming as reference the core level of the 4+ state, the V 2$p_{3/2}$ core level shifts by +4 eV for 2+ nominal valency, +2 eV for 3+, and --1 eV for 5+. The core levels in the initial state approximation (acceptable for localized perturbations)   are indeed consistent with the picture just outlined. For example, a single V in the presence of X, having acquired its two electrons, has M=3 and a shift of +4 eV consistent with expectations. This 2+ valency, however, is never going to materialize because V-X has a very high formation energy; this agrees with the absence of any XPS peak above the 3+ peak \cite{epfl}.

\begin{figure}[ht]
\centerline{\includegraphics[width=0.75\linewidth]{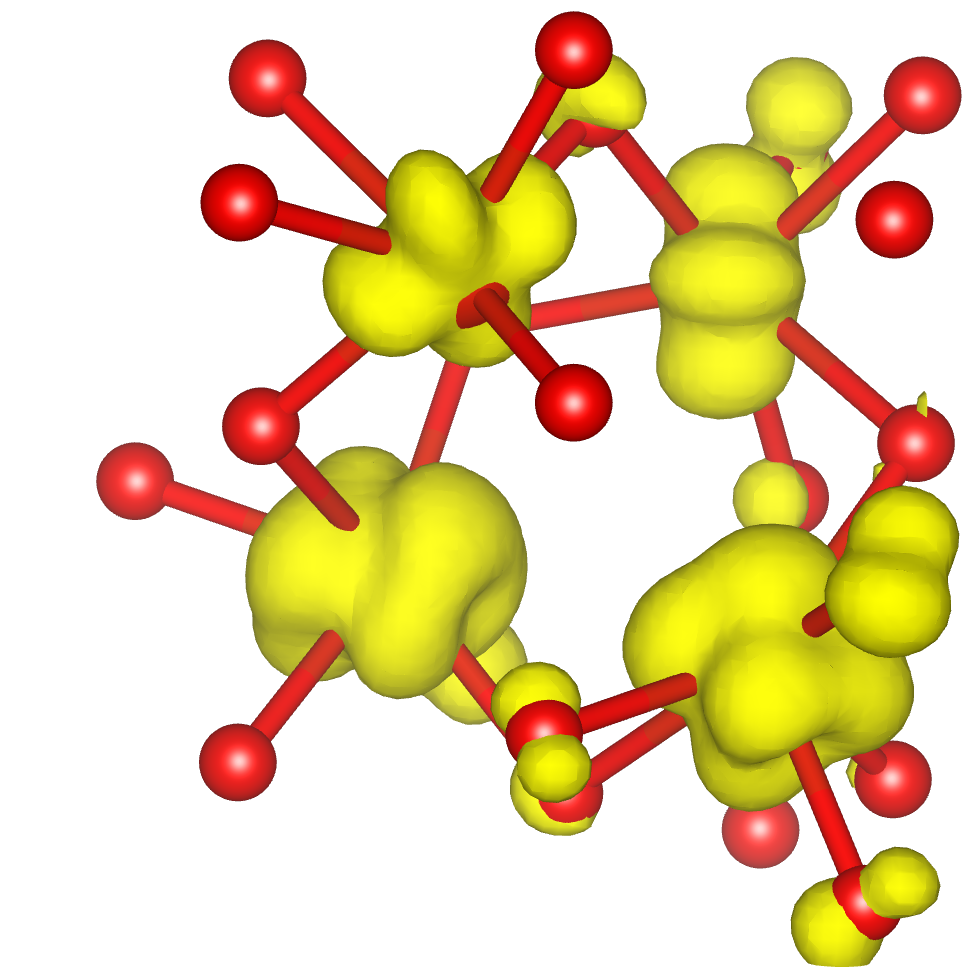}}
\caption{\label{magv4} Magnetization density of V$_4$ X (X is at the cluster center). The two lower V's have M=2, the upper ones M=1.  {Spheres are O atoms; V atoms are hidden by the magnetization cloud. The cluster is visually cut out of the periodic simulation cell for clarity}}
\end{figure}

In a V$_4$ with no vacancy each V  has  M=1, and indeed an average  shift of zero as expected; in the presence of a vacancy, two V have M=1.98 and accordingly  a +1.97 eV core-level shift, while the other two have M=1 and zero average shift. This is confirmed visually by the magnetization density in Figure \ref{magv4}.

Another interesting  example is V$_8$ X. In the parlance of this Section, one would expect six V's to be 4+ and two to be 3+ (those that acquire the two electrons from the vacancy). Local magnetizations and  core shifts show a slightly different picture: one of the  M=1, 4+ V's  releases its electron to another M=1, 4+ V, so that they end up with M=1.97 and a shift of 2.1 eV (i.e. a 3+ state) and, respectively, with M=0.04 and a shift of --1.1 eV, as expected for a 5+ valency. This  (relatively infrequent) occurrence accounts for the  weak XPS peak at the energy expected  for the 5+ ionicity \cite{epfl}. The main problem is of course the V${^3+}$ to V$^{4+}$ population ratio. This ratio,  discussed in Sec.\ref{ratio}, was zero by construction in \cite{io}, but was inferred to be around 2.8 experimentally from XPS signal decomposition \cite{stima,epfl}.

\subsection{Energetics and concentration of  V and X}
\label{energetics}

 Assuming no other donor is present, the 3+ to 4+ valency ratio is determined by the concentrations [V] and [X]  of vanadium and oxygen vacancies. [V] and [X] are calculated from the formation energies of variants of  V$_n$ X$_k$ embedded in hafnia. Entropy is neglected given its minor contributions suggested  by  the model of Ref.\cite{io} (as well as by a sample of direct calculations of the vibrational entropy). Only $k$=0, 1, and 2 are considered, as V$_n$X$_3$ is  found to be strongly  disfavored.

To obtain the energy of V$_n$X$_k$, an X is  added to  the V$_n$ configurations of Ref.\cite{io}. As these are spatially sparse for small $n$, possible compact V configurations near or around X are then searched for.  Such search is by necessity restricted to simple configurations obtained adding V atoms to compact clusters V$_n$X with small $n$ (1 to 4). Larger clusters are then  obtained adding extra V atoms to the outer border of the existing cluster. The V$_n$X$_2$ configurations are  obtained from the best V$_n$X adding a vacancy in a few selected O sites.  Despite the  uncertainties implicit in this approach, it is clear that compact V-X complexes are favored over sparse configurations.

\begin{figure}[ht]
\centerline{\includegraphics[width=0.8\linewidth]{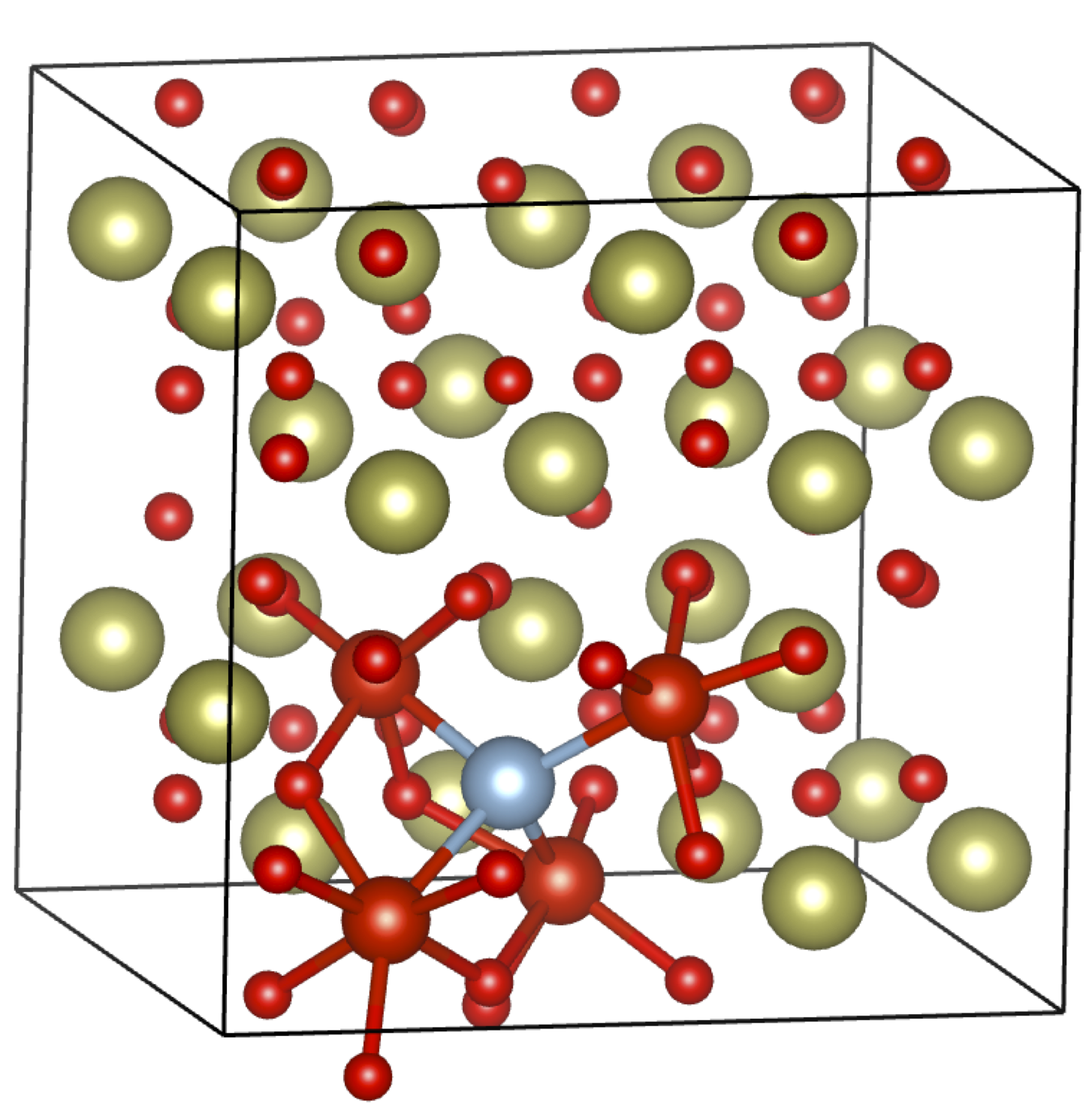}}
\caption{\label{V4X} An energetically-favored  compact V$_4$X cluster. {Color code for the atoms: large gold (light gray) is Hf, small red (dark gray) is O, medium red (dark gray) is V, medium light blue (light gray) is O vacant site.}}
\end{figure} 

V$_4$ X is a good example. An energy drop by 3.3 eV is found upon transforming an isolated X in hafnia plus a four-V supercell \cite{io} into X in the same four-V cell plus a bulk supercell. As discussed in Sec.\ref{gencons}, this $\sim$1.6 eV gain per electron is expected due to the transfer from X states high in the gap to V states low in the gap. 
Configuration sampling with X near a V, with X and two V neighbors, etc. eventually leads to the cluster in Figure \ref{V4X}, with an energy gain of 1.5 eV. This V$_4$X thus lowers the X formation energy by all of 4.8 eV compared to an isolated X in hafnia. Other V$_n$ X show a similar tendency to clustering around the vacancy. 

\begin{figure}[ht]
\centerline{\includegraphics[width=1\linewidth]{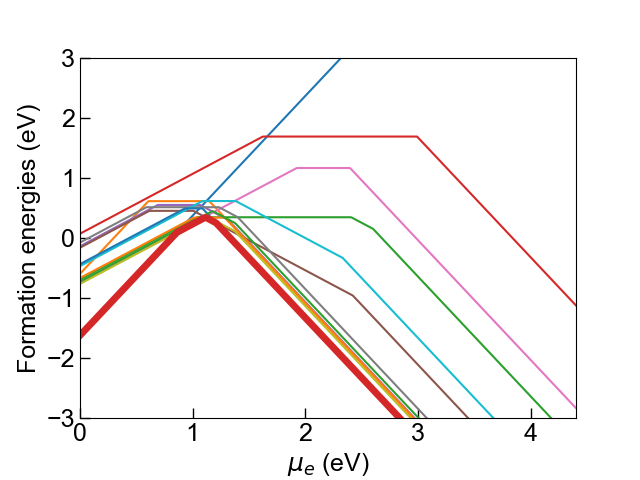}}
\caption{\label{efvsmue} Energies and lowest-energy envelope for V$_n$, X, and V$_n$ X$_{1,2}$ vs electron chemical potential $\mu_e$. Minimum (zero) $\mu_e$ is the valence band top, maximum is the Kohn-Sham gap. }
\end{figure}

Because V and V-X  clusters have available gap states, the chemical potential $\mu_e$ needs to be determined selfconsistently \cite{laks} from their charge states (from Q=+2 to --2, including standard finite-size corrections \cite{dE}). Figure \ref{efvsmue} reports energy  vs $\mu_e$ for various V-X, and their minimum-energy envelope, indicating a  generic self-compensating $p$-like behavior and  $\mu_e$$\sim$1.17 eV.

\begin{figure}[ht]
\centerline{\includegraphics[width=1\linewidth]{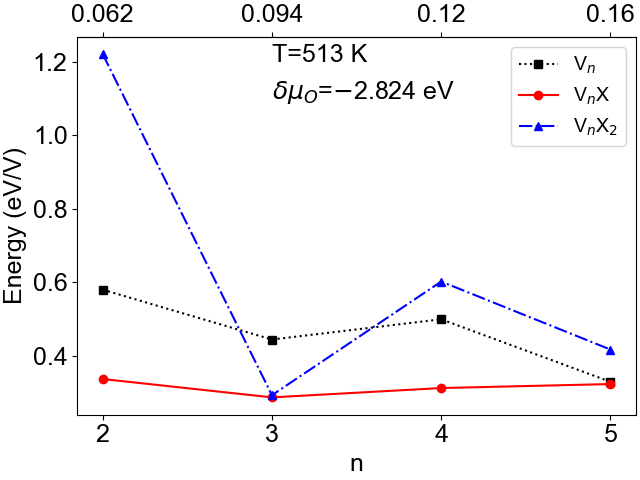}}
\caption{\label{Fclus} Formation energies of   V$_n$, V$_n$X, and V$_n$X$_2$ vs $n$. Top axis shows V concentration $x$=$n$/N$_{\rm cation}$=$n$/32. }
\end{figure}

Collecting all contributions, the energies as function of cluster size in Figure \ref{Fclus} are obtained. These energies depend 
on $\mu_e$, which is fixed as mentioned. They also depend, whenever X's are present, on the oxygen chemical potential $\mu_{\rm O}$. In Figure \ref{Fclus}, the value of $\mu_{\rm O}$ is the one reproducing (as discussed in the  next Section) the experimental 3+/4+ population ratio.

From these energies, the concentrations  [V] and [X] of V and X are obtained. Given the typical concentrations in experiment, and the predicted onset of phase separation at around 15\% \cite{io}, the sums making up [V] and [X] are limited to $n$$\leq$5 (although larger systems up to V$_{11}$X$_2$ were studied). Note that [V] increases more rapidly than [X] as $n$ grows, so hypothetical large clusters would suppress the 3+/4+ population ratio, which I now discuss.

\subsection{The 3+ to 4+ population ratio}
\label{ratio}

So far we have seen that each X transfers its electrons to two V, turning them from 4+ to 3+. As mentioned, the experimental ratio, by definition finite and non-negative, between V populations  with 3+ and 4+ valency is around 2.8 \cite{stima}. From our calculated concentrations, 
\begin{equation}
{\rm R}(T,\mu_{\rm O}) =\frac{[V^{\rm 3+}]}{[V^{\rm 4+}]}=\frac{[V^{\rm 3+}]}{[V] - [V^{\rm 3+}]}=\frac{2 [X]}{[V] - 2 [X]},
\label{Ratio}
\end{equation}
where the rare 5+ and 2+ states are implicitly neglected. R as just defined 
 is positive and finite if [X]$<$[V]/2. If [X]$\geq$[V]/2, all 4+ V disappear and R is no longer a meaningful indicator.

The energetics of X and V$_n$X$_k$ (and hence [X], [V], and R) depend directly on the chemical potential $\mu_{\rm O}$. If $\mu_{\rm O}$=$\mu_{\rm O}^{\rm mol}$$\equiv$$\mu_{\rm O_2}$/2 (energy per atom of the triplet O$_2$ molecule), the O thermodynamic reservoir is the molecular state, and the cost of oxygen deficiency is maximal. In such conditions, despite  the facilitating factor provided by electron transfer to V, the concentration [X] remains tiny. However, reducing conditions such that $\delta\mu_{\rm O}$=$\mu_{\rm O}$--$\mu_{\rm O}^{\rm mol}$$<$0 may enable larger [X]. Now $\delta\mu_{\rm O}$ is an externally determined  parameter that  varies between  zero and $\Delta H$/2$\simeq$--6 eV (half the formation enthalpy of hafnia). As shown below, R falls in the experimental range only for very reducing growth conditions, with $\delta\mu_{\rm O}$ about halfway the extremes just mentioned.

\begin{figure}[ht]
\centerline{\includegraphics[width=1\linewidth]{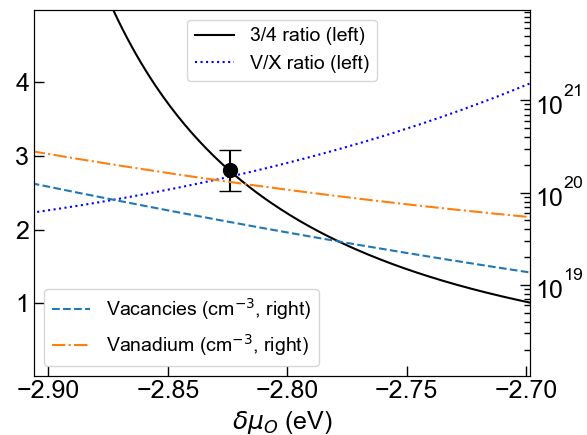}}
\caption{\label{R-mu} 
Ratio R vs. $\delta\mu_{\rm O}$ at T=513 K. The experimental R (black circle, error bar 10\%) is drawn in correspondence  to  $\delta\mu_{\rm O}$=--2.824 eV.}
\end{figure}

\begin{figure}[ht]
\centerline{\includegraphics[width=1\linewidth]{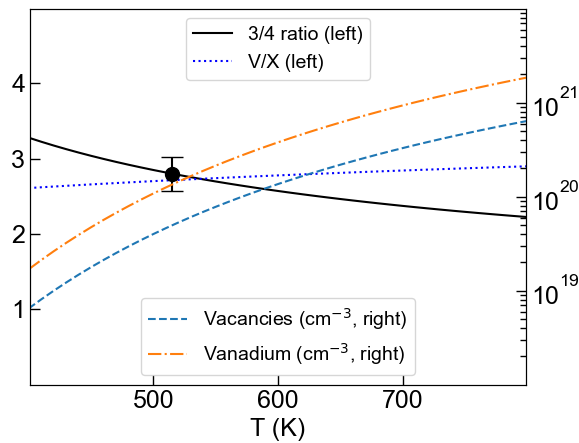}}
\caption{\label{R-T} Ratio R as function of T at $\delta\mu_{\rm O}$=--2.824 eV. Experimental R is the black circle, 10\% error bar.}
\end{figure} 

{Figure \ref{R-mu} displays R, [V], [X], and [V]/[X] vs  $\mu_{\rm O}$ at  the growth temperature T=513 K \protect\cite{epfl}.  Figure \ref{R-T} displays R, [V], [X], and [V]/[X] vs temperature, with $\delta\mu_{\rm O}$ fixed to the above value. The calculated ratio depends strongly on $\delta\mu_{\rm O}$ (note the narrow abscissa interval in the Figure). The $\delta\mu_{\rm O}$ reproducing the experimental R corresponds unambiguously to  strongly reducing conditions (its precise value, here --2.824 eV,  may depend on details and  technicalities).} 
 Due to its sensitivity to the chemical potential, the agreement with experiment would be spoiled  even for relatively mild deviations of $\delta\mu_{\rm O}$ away from this region.

So far we have seen that if X is assumed to be the only donor present, the observed R value can only occur for a restricted range of strongly  reducing growth conditions. 
However,  the chemistry of  atomic layer deposition (ALD) growth with precursors \cite{epfl} TEMAH, TEMAV, water and ozone is such 
that  there is no obvious built-in reducing agent in the gas phase \cite{ald}. For reasonable parameters, $\delta\mu_{\rm O}$ is indeed only around --0.1 eV, so quite O-rich {(and such that [X] is six orders of magnitude lower than needed).}

Such being the case, it is reasonable to conclude that either {\it a)} an unknown strongly reducing agent is present during growth, or {\it b)} the oxygen vacancies, while certainly present as described, are not the only source of electrons involved in the partial compensation of V from 4+ to 3+. This does not affect the prediction of an ``electronic reconstruction" V$^{4+}$+$e^-$$\rightarrow$V$^{3+}$ {\it per se}, but does raise questions about where the electrons actually come from.

It is true that, for plausible ALD water partial pressures 10$^{-2}$ to 1 mbar, even ppm-level H$_2$ partial pressures would suffice to bring $\mu_{\rm O}$ to the relevant range. On the other hand, a persistent  molecular hydrogen reservoir is unlikely because ozone is present in the growth sequence --indeed, in the very pulses introducing V-- and would efficiently oxidize H$_2$. Alternative or additional sources should involve extrinsic donors such as carbon/nitrogen residuals or such. This remains a puzzle, since the presence of sufficient concentrations  of donors other than oxygen vacancies was excluded \cite{epfl}.

{A side note about  vacancy migration: the diffusion barrier for charged vacancies (relevant here because vacancies cede their electrons to V) is 0.7 eV \cite{capron}. Thus, vacancies are effectively immobilized because they are  bound to V$_n$, which brings the effective barrier (escape from cluster plus diffusion) to 2-2.5 eV.}

\subsection{Implications for multiferroic Cr-doped hafnia}

The V-X results have implications for the recently reported multiferroicity in  Cr-doped hafnia. Unlike ALD-grown V:HfO$_2$ \cite{epfl}, Cr:HfO$_2$ is grown \cite{cr}  by spark plasma sintering, with precursors HfO$_2$ and Cr$_2$O$_3$ themselves, mixed so as to obtain  Cr concentrations around 10\%. Ideally, mixing a sesquioxide into a dioxide will cause \cite{cadel,eichler} one oxygen  vacancy to form for every two substituting ions, preserving their nominal 3+ ionicity. The XPS-measured 3+/4+ ratio is, however, finite and near 2, meaning that the reducing process is imperfect and a fraction of Cr's are 4+ rather than 3+. The discussion of the electronic structure then proceeds as for V  \cite{io}, with the electronic degeneracy removed by symmetry, and oxygen vacancy states higher in the hafnia gap releasing electrons to low-lying transition-metal states. 

The electrons provided by vacancies to the transition metal (V or Cr) will enhance the magnetization per site to the weighted average
\begin{equation}
{\rm M}_{\rm R} = \frac{{\rm M}^{3+} + {\rm M}^{4+}}{{\rm R}+1}=\frac{{\rm R} ({\rm M}+\delta{\rm M}) + {\rm M}}{{\rm R}+1}={\rm M} + \frac{\rm R\, \delta {\rm M}}{\rm R+1}
\nonumber\end{equation}
with M the value in the 4+ state, R the 3+/4+ ratio, and $\delta$M the magnetization change from 4+ to 3+. If each V or Cr collects only one additional electron (as is exceedingly likely), $\delta$M=1 $\mu_B$.

Since Cr has two native electrons in excess over Hf, if all spins are  ferromagnetically aligned, in the 4+ state its magnetization is M=2 $\mu_{\rm B}$/Cr, or  M$\sim$44 emu/cm$^3$ at 10\% Cr. Using R$\simeq$2 from XPS \cite{cr}, the enhanced magnetization M$_{\rm R}$ is 2.$\overline{6}$ $\mu_B$/Cr or 59 emu/cm$^3$ at 10\% Cr, very close to the measured \cite{cr} 60 emu/cm$^3$. {Also, the negligible magnetoanisotropy measured for Cr-doped hafnia \cite{cr} is compatible with that calculated for V-doped hafnia \cite{anis}}.
 Full Cr ferromagnetism as just assumed seems likely given the  Curie temperature in excess of 400 K \cite{cr}. 

{Note in passing that the calculated AF-FM difference of 0.02 meV/Cr in Ref.\cite{cr} is inconsistent with this important Curie temperature, which would require much larger values (46 meV in a cubic 3D Ising). 
Clearly a detailed analysis of  the Cr case is called for, and will be the subject of future work}.


The  V and Cr subplots of the multiferroic hafnia story are thus closely analogous, although distinct due to the respective growth conditions, and can be summarized as follows: ALD V:HfO$_2$ is globally O-rich, local ferromagnetic V and V-X complexes are expected, with a likely ``hidden" electron reservoir compounding  the effect of oxygen vacancies; SPS Cr:HfO$_2$ is genuinely oxygen-deficient, and  local ferromagnetic Cr or Cr-X complexes exist. In both cases, ferromagnetism originates from the peculiar symmetry of hafnia and the electronic structure of early transition metals, enhanced by oxygen vacancies (and/or hidden donors). 

While  the transition metal could cause magnetism by itself (as shown in \cite{io} for V), a useful side effect of X creation appears to be the improved stability \cite{cr,epfl} of ferroelectricity (and the increase in magnetization just reported). Incidentally, {ferroelectric polarization (obtained by the Berry-phase method \cite{berry}) remains largely unchanged as V$_n$X form (thus, presumably, Cr$_n$X as well); this should result in multiferroicity,} indeed seen experimentally for Cr:HfO$_2$.

\section{Summary and acknowledgments}

The interplay of oxygen deficiency and vanadium valency in V-doped  $Pca2_1$ hafnia HfO$_2$ was investigated ab initio. Oxygen vacancies donate electrons to low-lying majority V states, thereby lowering the vacancy formation energy, increasing the local magnetization, and reducing the nominal V ionicity from 4+ toward 3+. The calculated local moments and V core-level shifts consistently reproduce the experimental assignment of V$^{3+}$, V$^{4+}$, and weak V$^{5+}$
 components. However, reproducing the observed V$^{3+}$/V$^{4+}$
 ratio with oxygen vacancies alone requires considerably reducing conditions, suggesting that additional electron reservoirs may be active during ALD growth. Finally, the same vacancy-assisted valency reconstruction provides a natural explanation for the magnetization observed in Cr-doped  hafnia, where oxygen deficiency is expected from the synthesis route.
 
The author is grateful to Paola Alippi and Alessio Filippetti for comments; to RES-BSC Barcelona Supercomputing Center for resources on MareNostrum5 GPP; to CINECA Bologna for resources on EuroHPC-Leonardo; to TU Dresden for hosting him as a senior fellow during part of this work; to University of Cagliari for a sabbatical leave. Data supporting the findings  are  available \cite{dati}. 


\begin{thebibliography}{References}
\bibitem{ald} H$_2$ is not produced in the ligand exchange (byproducts are amines, not H$_2$ or hydrocarbons), and oxygen transitions directly from water into the solid.
\end{thebibliography}
\end{document}